# Magnetic anomalies in single crystalline $Tb_5Si_3$

**Kartik K Iyer,[1] K. Mukherjee,[1] P.L. Paulose,[1] E.V. Sampathkumaran,[1],\* Y. Xu,[2] and W. Löser[3]**

[1]Tata Institute of Fundamental Research, Homi Bhabha Road, Mumbai – 400005, India

[2]State Key Laboratory of Solidification Processing, Northwestern Polytechnical University, Xi'an, Shaanxi 710072, PRChina

[3]IFW Dresden, Leibniz-Institut für Festkörper- und Werkstoffforschung, Helmholtzstr.20, D-01171 Dresden, Germany

## ABSTRACT

The polycrystalline form of the compound, $Tb_5Si_3$, crystallizing in $Mn_5Si_3$-type hexagonal structure, which was earlier believe to order antiferromagnetically below 69 K, has been recently reported by us to exhibit interesting magnetoresistance (MR) anomalies. In order to understand the magnetic anomalies of this compound better, we synthesized single crystals of this compound and subjected to intense magnetization and MR studies. The results reveal that the magnetic behavior is strongly anisotropic as though the easy axis is along a basal plane. There appear to be multiple magnetic features in the close vicinity of 70 K. In addition, there are multiple steps in isothermal magnetization (which could not be resolved in the data for polycrystalline data) for magnetic-field ($H$) along a basal plane. The sign of MR is positive in the magnetically ordered state, and, interestingly, the magnitude dramatically increases at the initial step for $H$ parallel to basal plane, but decreases at subsequent steps as though the origin of these steps are different. However, for the perpendicular orientation ($H//[0001]$), there is no evidence for any step either in $M(H)$ or in MR($H$). These results establish this compound is an interesting magnetic material.

1. INTRODUCTION

It is a well-known fact that, across a metamagnetic transition, one observes a negative magnetoresistance, defined as MR= [$\rho(H)$-$\rho(0)$]/$\rho(0)$, where $\rho$ is the electrical resistivity. However, some few exceptions were known in the literature [see, for instance, Refs 1-4], in which cases the MR exhibits a jump at the positive side at the critical field ($H_c$) inducing the metamagnetic transition. Different explanations were proposed for this anomaly [1-4]. It is of interest to study such compounds in detail.

In this respect, the binary intermetallic compound, $Tb_5Si_3$ [5-7], crystallizing in $Mn_5Si_3$-type hexagonal structure (space group $P6_3/mcm$) is of special interest. The investigations on polycrystalline form revealed that there is a prominent antiferromagnetic ordering feature in this compound below 69 K. In the magnetically ordered state, there is a field-induced magnetic transition and $H_c$ increases with decreasing temperature in the forward leg of magnetic-field cycle, attaining a value of about 58 kOe at 1.8 K and there is a positive jump in MR at $H_c$ [3, 8-12]. A careful look at our results revealed that there could be more than one magnetic anomaly in the vicinity of 69 K in agreement with earlier literature [13] and that there could be additional steps in isothermal magnetization ($M$). We therefore considered it important to study the magnetization and electrical transport behavior of this compound in single crystalline form to understand this compound better.

2. EXPERIMENTAL DETAILS

The single crystal employed in the present investigation is the same as that in Ref. 14. Feed rods for crystal growth, 6mm in diameter and about 60 mm in length were prepared by melting together stoichiometric amounts of high-purity (>99.99%) Tb and Si in Hukin-type cold crucible induction furnace device. The crystal growth was performed with a floating-zone apparatus with optical radiation heating in a vertical double-ellipsoid optical configuration under 0.1 MPa flowing Ar atmosphere [14]. The $Tb_5Si_3$ single crystals of 6mm in diameter and 30 mm length were grown at 3mm/h zone travelling velocity. The crystals were characterized by x-ray diffraction and the lattice constants were found to be $a$= 8.469 Å and $c$= 6.354 Å. From the single crystal oriented by the x-ray Laue back scattering method, a right-angled block of 3×1.8×2 mm$^3$ size was prepared for measurements. The magnetic measurements were performed for two orientations, *H//[2 -1 -1 0]* and *H//[0001]* with the help of a commercial vibrating sample magnetometer (Quantum Design), as a function of temperature (*T*= 1.8 – 300 K) as well as of magnetic-field (up to 120 kOe). The ρ(*T, H*) studies were carried out using the Physical Properties Measurement System (Quantum Design) in the transverse mode. The data were collected after cooling the sample from 120 K to the desired temperature in zero magnetic field.

### 3. RESULTS AND DISCUSSIONS

The results of measurements as a function of temperature are shown in figures 1a and 1b for magnetization (measured in a field of 10 kOe) and in figures 1c and 1d for electrical resistvity (in zero field). It is clear that the properties are strongly anisotropic with larger values of these properties for *H* parallel to the basal plane, compared to the other orientation. It therefore appears that the moments point in the basal plane. A point of emphasis is that, for *H//[2 -1 -1 0]*, there is a peak in *M(T)* at 80 K as though there is a magnetic transition around this temperature. It should however be mentioned that traces (typically <1 %) of Tb$_5$Si$_4$, known to exhibit a magnetic transition near 80 K [15] could seen in the scanning electron microscopic pictures of single crystals [see figure 5b in Ref. 14]. Therefore, it is not clear whether this 80K-feature could arise from such a weak impurity phase. It is also known that this impurity appears only if the specimen is cooled slowly from the molten state - a procedure that is followed in the synthesis of single crystals. But this extra phase can be avoided if the cooling rate is fast, as evidenced by its absence in the arc-melted buttons of polycrystalline forms [14]. Though a feature attributable to 80K-transition could not be clearly visible in the ambient pressure data for polycrystalline sample, this could be resolved (even for polycrystals) if measured under a pressure of 10 kbar [8]. we therefore tend to believe that this 80K-transition could be intrinsic to the title compound. Turning to other features in the temperature dependence of magnetization, there is an upturn below 69 K, and sharp falls at 66 and 61 K. There are corresponding anomalies in the curve for H*//[0001]* and this curve further shows another feature at the tricritical point (near 20 K). A careful look at the $\rho(T)$ curves reveals that there is a sudden increase in the slope near 80 K, in addition to the drop at 66K (for both the orientations) and a shoulder near 60 K (for *H//[0001]*). Since there is no evidence for any other impurity phase containing Tb and Si (other than $Tb_5Si_4$), we conclude these additional features are intrinsic to $Tb_5Si_3$ phase.

We now turn to isothermal *M* behavior for *H//[2 -1 -1 0]* orientation, measured above 4.2 K, shown in figure 2. It may be recalled [3] that, in the case of polycrystals, there is a disorder-broadened field-induced first transitions near 58, 50, 50 and 30 kOe at 5, 10, 20 and 50 K respectively in the forward cycle with the hysteresis for temperatures below 20 K only. The transition in the reverse cycle for 5 and 10 K occurs near 28 and 40 kOe respectively. An inspection of the isothermal *M* data shown in figure 2 a-e reveals that there are additional transitions both in the forward as well as in the reverse cycles at all temperatures, apart from the fact that the first critical field values are marginally different from those for polycrystals. Thus, for instance, at 5 K, in addition to a jump around 55 kOe, another step is observed around 70-80 kOe in the forward cycle, whereas there is a broad feature near 55 kOe and a jump near 10 kOe in the reverse cycle. At 10 K, there are distinct transitions at 47 and 64 kOe in the forward cycle, and more jumps in the reverse cycle. Similar multiple transitions are visible at 20 K with hysteretic effect persisting at this temperature. Multiple transitions are seen even at a higher temperature, say near 24, 30, 34 and 40 kOe at 50 K. *M(H)* curves at 70 and 75 K look similar and the steps vanish as shown in figure 2e for 75 K, clearly revealing that the magnetic structure above and below 69 K are different.

With respect to the MR behavior for *H//[2 -1 -1 0]* orientation, though weak jumps seen in *M(H)* curves get smeared out, a careful look at the curves reveal that there are indeed multiple jumps or shoulders in this data as well at and below 50 K. A point to be noted is that, in the forward cycle, the electrical resistivity exhibits an upward jump at the initial transition, e.g., 55 kOe at 5 K, but at subsequent transitions, the resistivity interestingly decreases. This emphasizes that there could be a subtle difference in the physical origin of these transitions, for example, 'inverse metamagnetism' [3, 4, 8-10] near 55 kOe and well-known metamagnetism near 75 kOe for 5 K. Otherwise, other features (increase of MR in the reverse cycle with a peak and hysteretic effect) are qualitatively the same as that observed for polycrystalline samples and the readers may see earlier publications [3, 8-12] for the implications (e.g., unusual nature of magnetic phase co-existence phenomenon in some field-range while reversing the field, involving high-resistive high-field phase and low-resistive low-field phase). It is also to be noted that the value of MR at the peak of the curves is several hundred percent, much larger than that observed for polycrystals. We would like to emphasize that, above 69 K in a narrow vicinity of temperature, for instance at 75 K (Fig. 2 j), the behavior of MR is completely different in the sense that the sign of MR is negative and there is a sharp fall in magnitude, though weak, for initial applications of magnetic-field, followed by a more gradual variation with *H*. Since quadratic field dependence expected for paramagnetism at low-fields is not observed, the nature of the curve reveals persistence of magnetic ordering at such temperatures. Finally, a careful look at figure 2(f,g,h) shows that the transition-field in the forward cycle is rising as the temperature falls, but the one for the reverse leg is falling with temperature. Implication of this observation for phase-coexistence phenomenon as emphasized originally in Ref. 16 was already stressed for polycrystalline specimens [8] and similar finding was recently reported for $Er_5Si_3$ as well [17].

In the *M(H)* and MR data for *H//[0001]*, the behavior for H//*[2 -1 -1 0]* is completely different. We find that the jumps are absent. In the case of *M(H)* curves (see, for example, figure 3 for 5 and 50 K data), barring a weak change of slope for initial applications of magnetic-field, *M* varies essentially linearly with *H* (measured till 120 kOe). We attempted to measure MR as a function of magnetic field at several temperatures, but the sample gets twisted and detached from the sample holder due to the torque exerted by the magnetic-fields. We could however collect the data at 50 K till 60 kOe. We find that the sign of MR is positive with rather low magnitudes and it varies essentially quadratically with magnetic-field, indicating that what one observes for this orientation is the metallic contribution dominating over magnetic contribution.

## 4. SUMMARY

The magnetization and magnetoresistance studies on single crystals of $Tb_5Si_3$ suggest that this compound exhibits multiple magnetic transitions. Even as a function of magnetic-field, in the magnetically ordered state, there are multiple magnetic-field induced transitions for *H* along the basal plane. While the electrical resistivity at the initial critical-field inducing a magnetic transition increases, it decreases at the transitions occurring at higher fields, thereby implying that the origin of these transitions is different. It is thus interesting to see inverse-metamagnetic-like and metamagnetic-like behavior in the same material. However, such features are absent for *H//[0001]*. In addition, the magnitude of positive MR at the peak value while reversing the field is quite huge (several hundred percent at some temperatures). These results establish that this compound is an interesting magnetic material.

\*E-mail: sampath@mailhost.tifr.res.in

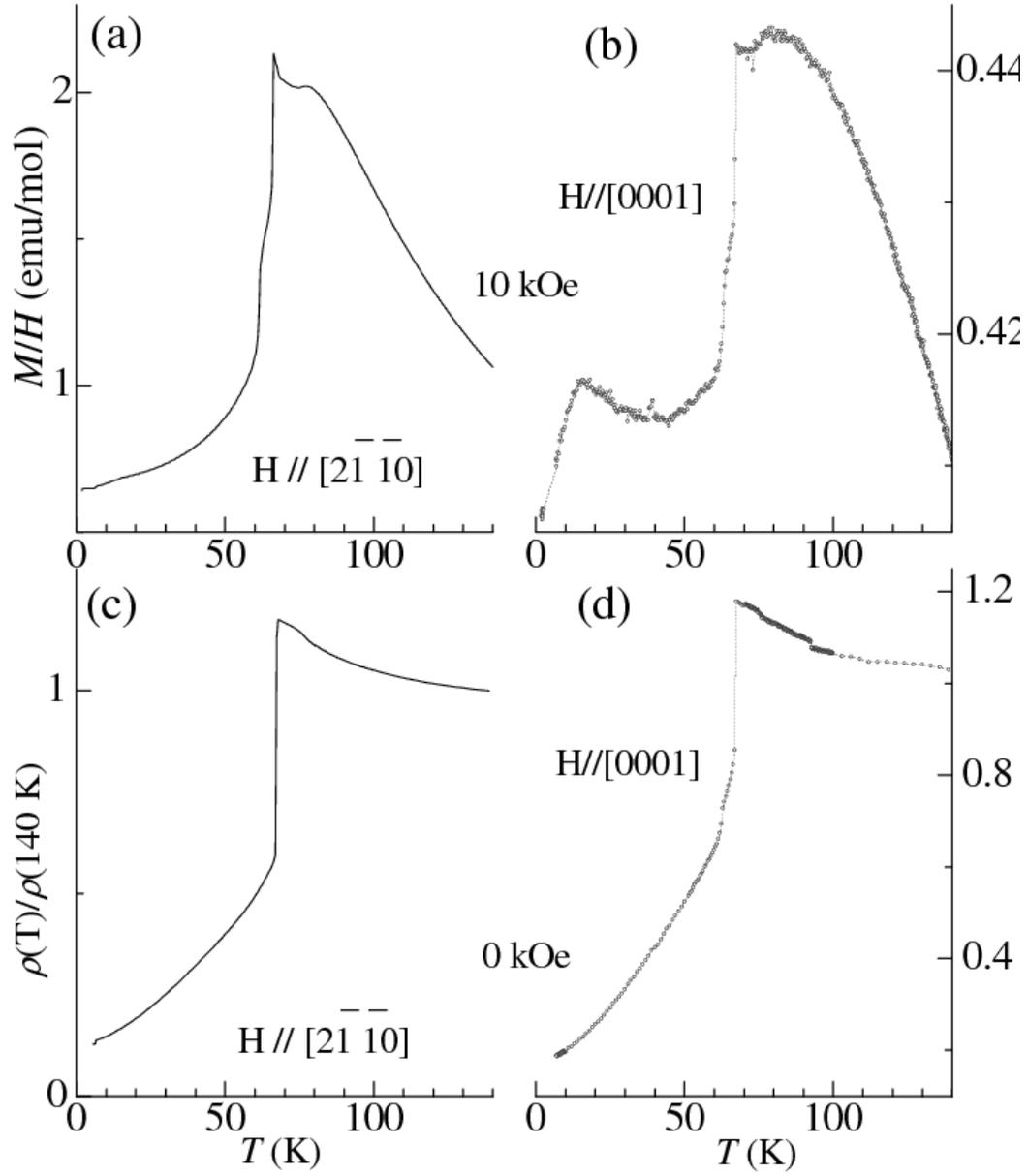

**Fig. 1:** Temperature dependence of magnetization along (a) $H//[2\ \text{-}1\ \text{-}1\ 0]$ and (b) $H//[0001]$ directions for single crystals of $Tb_5Si_3$. In (c) and (d), corresponding normalized electrical resistivity data are plotted.

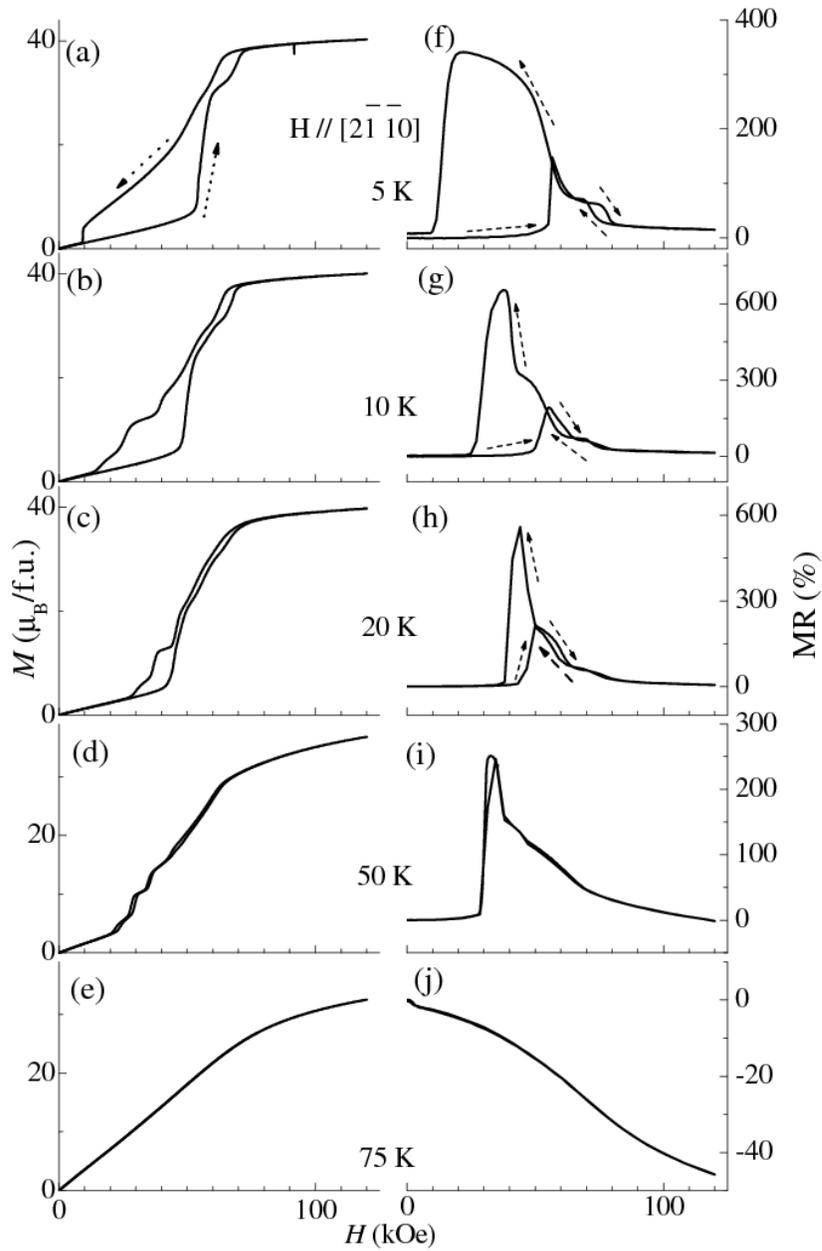

**Fig. 2:** Isothermal magnetization and magnetoresistance behavior of Tb$_5$Si$_3$ for *H//[2 -1 -1 0]* at various temperatures.

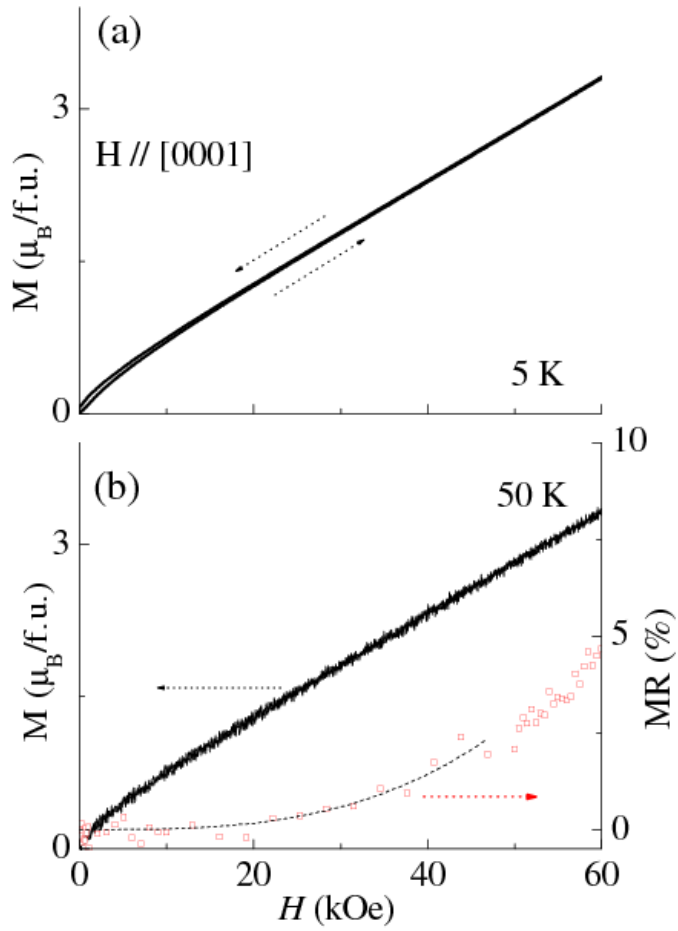

**Fig. 3:** Isothermal magnetization and magnetoresistance behavior of Tb$_5$Si$_3$ for *H//[0001]* at (a) 5 and (b) 50 K. (color online) Magnetoresistance behavior of Tb$_5$Si$_3$ for *H//[0001]* at 50 K. The dashed line shows quadratic fit to the data.